\documentclass[12pt]{article}
\usepackage{epsf}
\begin{document}

\title{Spectral Density of Complex Networks with Two Species of Nodes}

\author{Taro Nagao} 

\date{} 

\maketitle 

\begin{center} 
\it Graduate School of Mathematics,
Nagoya University, Chikusa-ku, \\ Nagoya 464-8602, Japan \\
\end{center}

\begin{abstract}
The adjacency and Laplacian matrices of complex networks with two species 
of nodes are studied and the spectral density is evaluated by using the 
replica method in statistical physics. The network nodes are classified 
into two species (A and B) and the connections are made only between 
the nodes of different species. A static model of such bipartite 
networks with power law degree distributions is introduced by applying 
Goh, Kahng and Kim's method to construct scale free networks. As a result, 
the spectral density is shown to obey a power law in the limit of 
large mean degree. 
\end{abstract}


\medskip


\newpage

\section{Introduction}
\setcounter{equation}{0}
\renewcommand{\theequation}{1.\arabic{equation}}

The theory of complex networks, which has dramatically 
been developed since the end of the last century, is based 
on the observation that there are universal features 
in real biological and social networks\cite{AB}. One of 
such features is the scale free property, meaning that the degree 
(the number of nodes directly connected to each node) 
distribution function $P(\Delta)$ obeys a power law $
P(\Delta) \propto \Delta^{-\lambda}$ for large 
$\Delta$. Barab\'asi and Albert explained the 
origin of this scale free property by focusing on the 
network growing process\cite{BA}. Goh, Kahng and Kim formulated 
a static network model which exhibits the scale free 
property\cite{GKK}.  
\par
The connection pattern of a network is mathematically 
described by the adjacency matrix. When the network 
has the scale free property, the spectral (eigenvalue) 
density $\rho(\mu)$ of the adjacency matrix is also 
expected to obey a power law $\rho(\mu) \propto 
\mu^{-\gamma}$ for large $\mu$. Dorogovtsev 
et al. presented an analytic evidence of this power 
law behaviour\cite{DGMS1,DGMS2}. Moreover a relation 
$\gamma = 2 \lambda - 1$ was found between the exponents 
of the power laws. Rodgers et al. analysed Goh, Kahng 
and Kim's static model and confirmed the power law 
behaviour of $\rho(\mu)$\cite{RAKK}. 
\par
In this paper, we shall study scale free networks 
with two species (A and B) of nodes. The connections 
are made only between the nodes of different species. 
We introduce a static model of such bipartite scale free 
networks by applying Goh, Kahng and Kim's method, and 
observe that each species has its own degree distribution 
function obeying a power law. Suppose that the exponent 
of the degree distribution function is $\lambda_A$ 
for the species A and $\lambda_B$ for the species B. 
Using the replica method in statistical physics, we 
are able to analytically evaluate the spectral density 
$\rho(\mu)$ in the limit of large mean degree\cite{RAKK,KK,RN}. 
As a result, we find that $\rho(\mu)$ also obeys a power law 
and the exponent $\gamma$ is associated with 
the exponents $\lambda_A$ and $\lambda_B$. In addition, 
the spectral density of the Laplacian matrix is similarly analysed 
and the power law behaviour is confirmed.
\par
Bipartite networks find applications in the analysis of 
human sexual contacts\cite{GE}, and of the connections 
between collaborators and collaboration acts\cite{RDPS}, 
such as actors and movies, scientists and papers. 
The adjacency matrices of bipartite networks are also 
interesting from the viewpoint of random matrix theory, 
since the Gaussian matrix model with the same structure 
is called the chiral Gaussian ensemble and applied to 
physics, such as the QCD gauge theory\cite{JV}.    
\par
The outline of this paper is as follows. In \S 2, 
a static model of bipartite scale free networks 
with two species of nodes is introduced, and the 
adjacency and Laplacian matrices are defined. 
In \S 3, in order to evaluate the spectral 
density, we apply the replica method to the 
network model. In \S 4, in the limit of large 
mean degree, the power law behaviour of the 
spectral density is analytically derived. 
In \S 4, the effective medium approximation 
is briefly discussed as an attempt to treat 
the case with a finite mean degree.

\section{Complex Networks with Two Species of Nodes}
\setcounter{equation}{0}
\renewcommand{\theequation}{2.\arabic{equation}}

Let us suppose that there are $N$ nodes of type A and 
$M$ nodes of type B ($N \geq M$). We are interested in 
the asymptotic behaviour of bipartite networks with two 
species of nodes A and B in the limit
\begin{equation}
\label{limit}
N \rightarrow \infty \ {\rm and} \ M \rightarrow \infty \ 
{\rm with} \ c = N/M \ {\rm fixed}.
\end{equation}
We introduce a static model of such networks with power law 
degree distributions by applying Goh, Kahng and Kim's method. 
Each node of type A is assigned a probability $P_j$ normalised as
\begin{equation}
\sum_{j=1}^N P_j = 1,
\end{equation}
while each node of type B has a probability $Q_k$ with
\begin{equation}
\sum_{k=1}^M Q_k = 1.
\end{equation}
The nodes of type A and B are connected according to the 
following procedure. In each step we choose a node $j$ of type A 
and a node $k$ of type B with probabilities $P_j$ and $Q_k$, 
respectively. Then the nodes $j$ and $k$ are connected, unless 
they are already connected. After repeating such a step $p N$ times, 
a node $j$ of type A and a node $k$ of type B is connected with 
a probability 
\begin{equation}
f_{jk} = 1 - (1 - P_j Q_k)^{p N} \sim 1 - {\rm e}^{- p N P_j Q_k}.
\end{equation}
\par
Let us consider an $N \times M$  matrix $C$ ($N \geq M$), where $C_{jk} 
= 1$ if the node $j$ of type A is directly connected to the node $k$ of 
type B, and $C_{jk} = 0$ otherwise. This random matrix $C$ describes 
the connection pattern of the network with two species of nodes. Eacn 
matrix element $C_{jk}$ is independently distributed with the probability 
density function (p.d.f.) 
\begin{equation}
\label{pjk}
{\cal P}_{jk}(C_{jk}) = (1 - f_{jk}) \delta(C_{jk}) 
+ f_{jk} \delta(1 - C_{jk}).
\end{equation}
We assume that $P_j$ and $Q_k$ are given by 
\begin{equation}
\label{pj}
P_j = \frac{j^{-\alpha}}{\displaystyle 
\sum_{l=1}^N l^{-\alpha}} \sim (1 - \alpha) N^{\alpha - 1} 
j^{-\alpha}, \ \ \ 0 < \alpha < 1
\end{equation}
and
\begin{equation}
Q_k = \frac{k^{-\beta}}{\displaystyle 
\sum_{l=1}^M l^{-\beta}} \sim (1 - \beta) M^{\beta - 1} 
k^{-\beta}, \ \ \ 0 < \beta < 1.
\end{equation}
There are thus two parameters $\alpha$ and $\beta$ controlling 
the p.d.f. of the matrix $C$.
\par 
We define the degree $d_j$ of the type-A node $j$ 
as the number of directly connected type-B nodes: 
\begin{equation}
d_j = \sum_{k=1}^M C_{jk}.
\end{equation}
Then the type-A node degree distribution function is given by 
\begin{equation}
\label{padelta}
P^{(A)}(\Delta) = \left\langle \frac{1}{N} 
\sum_{j=1}^N \delta(\Delta - d_j) \right\rangle,
\end{equation}
where the brackets denote the average over the p.d.f. (\ref{pjk}) 
and $\delta(x)$ is Dirac's delta function. We can similarly 
introduce the degree $e_k$ of the type-B node $k$: 
\begin{equation}
e_k = \sum_{j=1}^N C_{jk}
\end{equation}
and the type-B node degree distribution function
\begin{equation}
P^{(B)}(\Delta) = \left\langle \frac{1}{M} 
\sum_{k=1}^M \delta(\Delta - e_k) \right\rangle.
\end{equation}
\par
In Appendix A, a useful asymptotic relation 
\begin{equation}
\label{asymptotic} 
\ln \left\langle {\rm exp}\left( - i \sum_{j=1}^N \sum_{k=1}^M 
C_{jk} t_{jk} \right) \right\rangle \sim 
p N \sum_{j=1}^N \sum_{k=1}^M P_j Q_k({\rm e}^{ - i t_{jk}} - 1)
\end{equation}
is derived in the limit (\ref{limit}). Here $t_{jk}$, which 
depends on neither $N$ nor $M$, is in the neighbourhood of 
the origin so that $|{\rm e}^{ - i t_{jk}} - 1| < 1$.
\par 
As special cases, we can readily derive asymptotic relations for
\begin{equation}
F^{(A)}_j(t) \mathrel{\mathop:}= \ln \left\langle {\rm e}^{- i d_j t} \right\rangle, \ \ \   
F^{(B)}_k(t) \mathrel{\mathop:}= \ln \left\langle {\rm e}^{- i e_k t} \right\rangle 
\end{equation}
as 
\begin{equation}
\label{fatfbt}
F^{(A)}_j(t) \sim p N P_j ({\rm e}^{- i t} - 1), \ \ \  
F^{(B)}_k(t) \sim p N Q_k ({\rm e}^{- i t} - 1). 
\end{equation} 
Then we can readily see that 
\begin{eqnarray}
\langle d_j \rangle & = & i \left. 
\frac{\partial}{\partial t} F^{(A)}_j(t) \right|_{t=0} 
\sim p N P_j, \nonumber \\  
\langle e_k \rangle & = & i \left. 
\frac{\partial}{\partial t} F^{(B)}_k(t) \right|_{t=0} 
\sim p N Q_k,   
\end{eqnarray}
so that the mean degree $m^{(A)}$ of the type-A node is
\begin{equation}
m^{(A)} = \frac{1}{N} \sum_{j=1}^N \langle d_j \rangle 
\sim p,
\end{equation}
while the mean degree $m^{(B)}$ of the type-B node is
\begin{equation}
m^{(B)} = \frac{1}{M} \sum_{k=1}^M \langle e_k \rangle 
\sim p c.
\end{equation}
\par
It can be seen from (\ref{pj}), (\ref{padelta}) and (\ref{fatfbt}) that 
the type-A node degree distribution function can be written as
\begin{eqnarray}
P^{(A)}(\Delta) & = & \frac{1}{2 \pi N} \sum_{j=1}^N \int {\rm d}t 
 \ {\rm e}^{i \Delta t + F^{(A)}_j(t)} \nonumber \\ 
& \sim & \frac{1}{2 \pi} \int {\rm d}t \int_0^1 {\rm d}x 
 \ {\rm exp}\left\{ i \Delta t + p (1 - \alpha) x^{-\alpha} 
({\rm e}^{- i t} - 1) \right\}.
\end{eqnarray}
Then in the limit $\Delta \rightarrow \infty$ we find
\begin{equation}
P^{(A)}(\Delta) \sim \int_0^1 {\rm d}x \ \delta\left\{ 
\Delta - p (1 - \alpha) x^{-\alpha} \right\} 
= \frac{\{ p (1 - \alpha) \}^{1/\alpha}}{\alpha} 
\frac{1}{\Delta^{(1/\alpha) + 1}},
\end{equation}
and similarly obtain  
\begin{equation}
P^{(B)}(\Delta) \sim \int_0^1 {\rm d}y \ \delta\left\{ 
\Delta - p c (1 - \beta) y^{-\beta} \right\} 
= \frac{\{ p c (1 - \beta) \}^{1/\beta}}{\beta} 
\frac{1}{\Delta^{(1/\beta) + 1}}.
\end{equation}
Thus we have seen that the network has the scale free property, 
as the node degree distribution functions obey power laws. 
The exponents of the power laws defined as
\begin{equation}
P^{(A)}(\Delta) \propto \Delta^{-\lambda_A}, \ \ \ 
P^{(B)}(\Delta) \propto \Delta^{-\lambda_B}, \ \ \ \Delta \rightarrow \infty
\end{equation}
are found to be $\lambda_A = (1/\alpha)+1$ and $\lambda_B = (1/\beta)+1$. 
\par 
In this paper we study the adjacency and Laplacian matrices of 
this scale free network. The adjacency matrix ${\cal A}$ of this 
network is defined as 
\begin{equation}
{\cal A} = \left( \begin{array}{cc} O_N & C \\ 
C^{\rm T} & O_M \end{array} \right),
\end{equation}
where $C^{\rm T}$ is the transpose of $C$ and $O_n$ is an $n \times n$ 
matrix with zero elements. The Laplacian matrix ${\cal L}$ is an $(N + M) 
\times (N + M)$ symmetric matrix with 
\begin{equation}
{\cal L}_{jl} = \left\{ \begin{array}{ll} 
d_j, & j = l \ {\rm and} \ 1 \leq j \leq N, \\    
e_{j-N}, & j = l \ {\rm and} \ N+1 \leq j \leq N+M, \\ 
- {\cal A}_{jl}, & j \neq l. \end{array} \right.
\end{equation}

\section{Spectral Density}
\setcounter{equation}{0}
\renewcommand{\theequation}{3.\arabic{equation}}

Let us define that $J$ is the adjacency matrix ${\cal A}$ or the 
Laplacian matrix ${\cal L}$. The spectral density of $J$ 
is defined as 
\begin{equation}
\rho(\mu) = \left\langle \frac{1}{N+M} \sum_{j=1}^{N+M} \delta(\mu - \mu_j) 
\right\rangle,
\end{equation}
where $\mu_j$, $j = 1,2,\cdots,N+M$ are the eigenvalues of $J$. 
In order to calculate $\rho(\mu)$, we introduce the partition function  
\begin{equation}
\label{pf}
Z(\mu) = \int^{\infty}_{-\infty} \prod_{j=1}^{N+M} {\rm d}\Phi_j \ {\rm exp}\left( \frac{i}{2} \mu 
\sum_{j=1}^{N+M} \Phi_j^2 - \frac{i}{2} \sum_{j=1}^{N+M} \sum_{l=1}^{N+M} J_{jl} 
\Phi_j \Phi_l \right).
\end{equation}
Using the partition function $Z$, we can write the spectral density as
\begin{eqnarray}
\rho(\mu) & = & \frac{1}{(N+M) \pi} {\rm Im}{\rm Tr}\left\langle 
\{J - (\mu + i \epsilon) I\}^{-1} \right\rangle \nonumber \\ 
& =  & \frac{2}{(N+M) \pi} {\rm Im} \frac{\partial}{\partial \mu} 
\langle \ln Z(\mu + i \epsilon) \rangle,
\end{eqnarray}      
where $\epsilon$ is an infinitesimal positive number and $I$ is an 
$(N+M) \times (N+M)$ identity matrix. Then we can utilise the relation  
\begin{equation}
\lim_{n \rightarrow 0} \frac{\ln \langle Z^n \rangle}{n} 
= \langle \ln Z \rangle
\end{equation}
to obtain
\begin{equation}
\label{rhomu}
\rho(\mu) = \lim_{n \rightarrow 0} \frac{2}{(N+M) n \pi} {\rm Im} 
\frac{\partial}{\partial \mu} 
\ln \langle \{ Z(\mu) \}^n \rangle.
\end{equation}
Therefore it is necessary to evaluate the average $\langle Z^n \rangle$. 
\par
The replica method explained in Appendix B is known to be a powerful 
tool for that purpose. It follows in the limit (\ref{limit}) that  
\begin{equation}
\label{s0s1s2}
\langle Z^n \rangle \sim \int 
\prod_{j=1}^N {\cal D}\xi_j({\vec \phi}) 
\prod_{k=1}^M {\cal D}\eta_k({\vec \psi}) 
\ {\rm e}^{S_0 + S_1 + S_2}.
\end{equation}
Here
\begin{equation}
\label{s0}
S_0 = - \sum_{j=1}^N \int {\rm d}{\vec \phi} 
\ \xi_j({\vec \phi}) \ln \xi_j({\vec \phi}) 
- \sum_{k=1}^M \int {\rm d}{\vec \psi} 
\ \eta_k({\vec \psi}) \ln \eta_k({\vec \psi}),
\end{equation}
\begin{equation}
\label{s1}
S_1 = \frac{i}{2} \mu \sum_{j=1}^N \int {\rm d}{\vec \phi} 
\ \xi_j({\vec \phi}) {\vec \phi}^2 + 
\frac{i}{2} \mu \sum_{k=1}^M \int {\rm d}{\vec \psi} 
\ \eta_k({\vec \psi}) {\vec \psi}^2
\end{equation}
and
\begin{equation}
\label{s2}
S_2 = p N \sum_{j=1}^N \sum_{k=1}^M P_j Q_k 
\int {\rm d}{\vec \phi} \int {\rm d}{\vec \psi} 
\ \xi_j({\vec \phi}) \ \eta_k({\vec \psi}) \left\{ 
f({\vec \phi},{\vec \psi}) - 1 \right\}
\end{equation}
with
\begin{equation}
\label{fpp}
f({\vec \phi},{\vec \psi})
= \left\{ \begin{array}{ll} 
{\rm e}^{- i {\vec \phi} \cdot {\vec \psi}}, 
& {\rm if} \ J \ {\rm is} \ {\rm the} \ {\rm adjacency} 
\ {\rm matrix} \ {\cal A}, \\ 
{\rm e}^{- \frac{i}{2}({\vec \phi} -{\vec \psi})^2}, 
& {\rm if} \ J \ {\rm is} \ {\rm the} \ {\rm Laplacian} 
\ {\rm matrix} \ {\cal L}. 
\end{array} \right.
\end{equation}
The functional integrations are taken over the auxiliary functions 
$\xi_j({\vec \phi})$ and $\eta_k({\vec \psi})$ satisfying 
\begin{equation}
\label{xieta}
\int {\rm d}{\vec \phi} \ \xi_j({\vec \phi}) = 
\int {\rm d}{\vec \psi} \ \eta_k({\vec \psi}) = 1.
\end{equation}
\par
In the limit (\ref{limit}), the functional integrations over 
$\xi_j({\vec \phi})$ and $\eta_k({\vec \psi})$ are dominated 
by the stationary point satisfying
\begin{equation}
\delta\left\{ 
S_0 + S_1 + S_2 
+ \sum_{j=1}^N \theta_j \left( \int {\rm d}{\vec \phi} \ \xi_j({\vec \phi}) - 1 \right) 
+ \sum_{k=1}^M \omega_k \left( \int {\rm d}{\vec \psi} \ \eta_k({\vec \psi}) - 1 \right) 
\right\} = 0,
\end{equation} 
where $\theta_j$ and $\omega_k$ are the Lagrange multipliers. It follows from 
this equation that
\begin{eqnarray}
\label{variational}
\xi_j({\vec \phi}) & = & \Theta_j \ {\rm exp}\left[ \frac{i}{2} \mu {\vec \phi}^2 + 
p N P_j \sum_{k=1}^M Q_k \int {\rm d}{\vec \psi} \eta_k({\vec \psi}) 
\left\{ f({\vec \phi},{\vec \psi}) - 1 \right\} \right], \nonumber \\ 
\eta_k({\vec \psi}) & = & \Omega_k \ {\rm exp}\left[ \frac{i}{2} \mu {\vec \psi}^2 + 
p N Q_k \sum_{j=1}^N P_j \int {\rm d}{\vec \phi} \xi_j({\vec \phi}) 
\left\{ f({\vec \phi},{\vec \psi}) - 1 \right\} \right], \nonumber \\  
\end{eqnarray}
where $\Theta_j$ and $\Omega_k$ are normalisation constants.
\par
In the limit of large mean degree $p \rightarrow \infty$, 
the variational equations (\ref{variational}) are satisfied 
by the Gaussian ansatz 
\begin{equation}
\label{gaussian}
\xi_j({\vec \phi}) = \frac{1}{(2 \pi i \sigma_j)^{n/2}} \ 
{\rm exp} \left(  - \frac{{\vec \phi}^2}{2 i \sigma_j} \right), \ \ \   
\eta_k({\vec \psi}) = \frac{1}{(2 \pi i \tau_k)^{n/2}} \ 
{\rm exp} \left(  - \frac{{\vec \psi}^2}{2 i \tau_k} \right),   
\end{equation}
as shown in Appendix C. Here ${\rm Im} \sigma_j \leq 0$ and ${\rm Im} 
\tau_k \leq 0$. This property simplifies the problem and enables us 
to evaluate the asymptotic spectral density.
\par
Let us consider the limit $p \rightarrow \infty$ with a scaling variable 
$E = \mu/\sqrt{p}$. In Appendix C, we find the asymptotic spectral 
density of the adjacency matrix ${\cal A}$ as
\begin{equation}
\rho(\mu) \sim \frac{2}{\sqrt{p} (1 + c)} \left\{ 
\frac{c^{1/\beta} (1 - \beta)^{1/\beta}}{\beta E^{(2/\beta)+ 1}}   
+ \frac{c (1 - \alpha)^{1/\alpha}}{\alpha E^{(2/\alpha)+ 1}} \right\}
\end{equation}
in the tail region $E \rightarrow \infty$. The exponent $\gamma$ of 
the spectral density defined as
\begin{equation}
\rho(\mu) \propto \mu^{-\gamma}, \ \ \ \mu \rightarrow \infty
\end{equation}
is $(2/\alpha) + 1$ if $\alpha \geq \beta$, and is $(2/\beta) + 1$ 
if $\beta \geq \alpha$. Thus $\gamma$ is associated with 
$\lambda_A = (1/\alpha) + 1 $ and $\lambda_B = (1/\beta) + 1$ 
as $\gamma = 2 \min(\lambda_A,\lambda_B)-1$. 
\par
It is also explained in Appendix C that the asymptotic spectral density 
of the Laplacian matrix ${\cal L}$ is given by
\begin{eqnarray}
\rho(\mu) & \sim & 
\frac{c \{p (1 - \alpha)\}^{1/\alpha}}{(1 + c) \alpha} 
\frac{1}{\mu^{(1/\alpha)+1}} \ H\left\{ \mu - p (1 - \alpha) \right\}  
\nonumber \\ 
& & + \frac{\{p c (1 - \beta)\}^{1/\beta}}{(1 + c) \beta} 
\frac{1}{\mu^{(1/\beta)+1}} \ H\left\{ \mu - p c (1 - \beta) \right\}
\end{eqnarray}
in the region $\mu = O(p)$ with $p \rightarrow \infty$. Here 
$H(x)$ is defined as
\begin{equation}
\label{hx}
H(x) = \left\{ \begin{array}{ll} 0, & x < 0, \\ 
1, & x > 0. \end{array} \right.
\end{equation} 
The exponent $\gamma_L$ of the spectral density $\rho(\mu) \propto \mu^{-\gamma_L}$ 
($\mu \rightarrow \infty$) is $(1/\alpha) + 1$ if $\alpha \geq \beta$, 
and is $(1/\beta) + 1$ if $\beta \geq \alpha$. Thus $\gamma_L$ is 
associated with $\lambda_A$ and $\lambda_B$ as $\gamma_L = 
\min(\lambda_A,\lambda_B)$. 

\section{Effective Medium Approximation}
\setcounter{equation}{0}
\renewcommand{\theequation}{4.\arabic{equation}}

In the previous section we have dealt with the spectral density 
in the limit $p \rightarrow \infty$. The calculation of the spectral 
density with a finite mean degree $p$ is a much more involved problem, 
for which sophisticated numerical schemes have been 
proposed\cite{RK,RCKT,EK,KM}. In this section we briefly discuss 
a simple approximation method (effective medium approximation) 
for that problem\cite{RN,RB,BR,SC,NT,NR}. In this approximation, 
we put the Gaussian ansatz (\ref{gaussian}) into the formulas 
(\ref{s0}), (\ref{s1}) and (\ref{s2}), and solve the stationary 
point equations
\begin{equation}
\frac{\partial}{\partial \sigma_j} (S_0 + S_1 + S_2) = 0
\end{equation}
and
\begin{equation}
\frac{\partial}{\partial \tau_k} (S_0 + S_1 + S_2) = 0.
\end{equation}
\par
In the case of the adjacency matrix ${\cal A}$, the above procedure results 
in the effective medium approximation (EMA) equations
\begin{eqnarray}
\label{ema}
& & \mu - \frac{1}{\sigma_j} - p N P_j \sum_{k=1}^M \frac{Q_k 
\tau_k}{1 - \sigma_j \tau_k} = 0,
\nonumber \\ 
& & \mu - \frac{1}{\tau_k} - p N Q_k \sum_{j=1}^N \frac{P_j 
\sigma_j}{1 - \sigma_j \tau_k} = 0.
\end{eqnarray}
As for scale free networks with a single species of nodes, 
Nagao and Rodgers calculated the $1/p$ expansion of the spectral 
density by using the corresponding EMA equation\cite{NR}. A similar analytical 
treatment could also be possible in the present case. Here, however 
only results of numerical iterations of (\ref{ema}) are shown in 
Figure 1 as the EMA spectral densities. They are compared with the 
spectral densities of positive eigenvalues calculated by numerical 
diagonalisations of numerically generated adjacency matrices 
(averaged over $100$ samples). The EMA gives a better fit for a 
larger $p$, as expected from the fact that the variational equations 
(\ref{variational}) are satisfied by the Gaussian ansatz (\ref{gaussian}) 
in the limit $p \rightarrow \infty$. When $p = 1$, the agreement 
significantly breaks down around the origin, although it is still 
fairly good in the tail region with large $\mu$.   

\begin{figure}[ht]
\epsfxsize=14cm
\centerline{\epsfbox{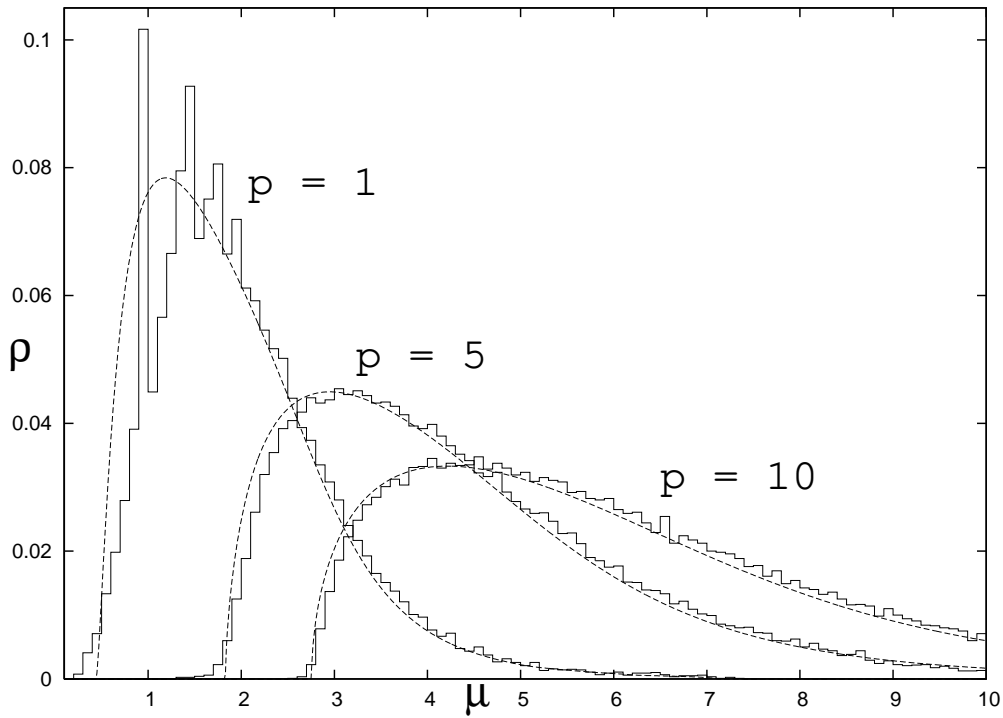}}
\caption{The EMA spectral densities (dashed curves) and the spectral 
densities of numerically generated adjacency matrices (histograms) 
with $p = 1,5$ and $10$. The parameters are $N=1000$, $M=200$ and 
$\alpha = \beta = 1/2$.} 
\end{figure}

In the limit $\alpha, \beta \rightarrow 0$, we obtain the adjacency 
matrix of a classical random graph with two species, where the 
connections are made only between the nodes of different species. 
In that case $\sigma_j$ and $\tau_k$ can be written as $\sigma$ and 
$\tau$, respectively, because they depend on neither $j$ nor $k$. 
The EMA equations become a cubic equation for $\sigma$ 
\begin{equation}
c \sigma^3 + \frac{c (p - 2) + 1}{\mu} \sigma^2 - 
\frac{\mu^2  + (c - 1)(p - 1)}{\mu^2} \sigma + \frac{1}{\mu} = 0 
\end{equation}
and
\begin{equation}
\tau = \frac{c (\mu \sigma - 1) + 1}{\mu}.
\end{equation}
These equations are equivalent to Nagao and Tanaka's SEMA (symmetric 
EMA) equations concerning the spectral density of sparse correlation 
matrices\cite{NT}, and can be analysed in the same way.
\par 
We can similarly derive the EMA equations for the Laplacian matrix 
${\cal L}$ as 
\begin{eqnarray}
& & \mu - \frac{1}{\sigma_j} - p N P_j \sum_{k=1}^M \frac{Q_k}{1 
- \sigma_j - \tau_k} = 0,
\nonumber \\ 
& & \mu - \frac{1}{\tau_k} - p N Q_k \sum_{j=1}^N \frac{P_j}{1 
- \sigma_j - \tau_k} = 0.
\end{eqnarray}
In the limit $\alpha, \beta \rightarrow 0$, $\sigma_j$ and $\tau_k$ 
can again be reduced to $\sigma$ and $\tau$, respectively. Then we 
find a cubic equation for $\sigma$ 
\begin{equation} 
(1 - c) \sigma^3 + \frac{2 c + (p - \mu)(1 - c)}{\mu} \sigma^2 + 
\frac{\mu - 1 + c (p - 2 \mu - 1)}{\mu^2} \sigma + \frac{c}{\mu^2} = 0 
\end{equation}
and
\begin{equation}
\tau = \frac{\sigma}{(1 - c)\mu \sigma + c}.
\end{equation}

\section*{Acknowledgements}

The author thanks Prof. G.J. Rodgers and Prof. Toshiyuki Tanaka for valuable 
discussions. This work was partially supported by the Japan Society for the 
Promotion of Science (KAKENHI 20540372).

\section*{Appendix A}
\setcounter{equation}{0}
\renewcommand{\theequation}{A.\arabic{equation}}

In this Appendix, we derive an asymptotic relation
\begin{equation} 
\ln \left\langle {\rm exp}\left( - i \sum_{j=1}^N \sum_{k=1}^M 
C_{jk} t_{jk} \right) \right\rangle \sim 
p N \sum_{j=1}^N \sum_{k=1}^M P_j Q_k ({\rm e}^{ - i t_{jk}} - 1),
\end{equation}
where $t_{jk}$ is a parameter which is independent of $N$ and $M$. 
We moreover assume that $t_{jk}$ is in the neighbourhood of the 
origin so that $|S_{jk}| < 1$ holds for $S_{jk} = {\rm e}^{ - i t_{jk}} - 1$. 
A similar argument for Goh, Kahng and Kim's model is 
found in \cite{KRKK}. 
\par
The Taylor expansion of the logarithmic function gives
\begin{eqnarray}
& & \ln \left\langle {\rm exp}\left( - i \sum_{j=1}^N \sum_{k=1}^M 
C_{jk} t_{jk} \right) \right\rangle = \sum_{j=1}^N \sum_{k=1}^M 
\ln(1 + f_{jk} S_{jk}) \nonumber \\ 
& = & p N \sum_{j=1}^N \sum_{k=1}^M P_j Q_k S_{jk} + 
\sum_{j=1}^N \sum_{k=1}^M (f_{jk} - p N P_j Q_k) S_{jk} 
\nonumber \\ & & + \sum_{\ell=2}^{\infty} \frac{(-1)^{\ell+1}}{\ell} 
\sum_{j=1}^N \sum_{k=1}^M (f_{jk} S_{jk})^\ell.
\end{eqnarray}
We show
\begin{equation}
\ln \left\langle {\rm exp}\left( - i \sum_{j=1}^N \sum_{k=1}^M 
C_{jk} t_{jk} \right) \right\rangle - p N \sum_{j=1}^N \sum_{k=1}^M 
P_j Q_k S_{jk} = o(N)  
\end{equation}
in two steps.  
\par
\medskip
\noindent
{\bf Step 1} 
\par
\medskip
\noindent
Let us first prove that 
\begin{equation}
\label{2nd}
\sum_{j=1}^N \sum_{k=1}^M (f_{jk} - p N P_j Q_k) S_{jk} = o(N).
\end{equation} 
We define
\begin{equation}
S_{\max} = \max_{jk} | S_{jk} |
\end{equation}
and
\begin{equation}
G_1(x) = x - 1 + {\rm e}^{-x}.
\end{equation}
Then we see that    
\begin{eqnarray}
\left| \sum_{j=1}^N \sum_{k=1}^M (f_{jk} - p N P_j Q_k) S_{jk} \right| 
& \leq & S_{\max} \sum_{j=1}^N \sum_{k=1}^M |f_{jk} - p N P_j Q_k| 
\nonumber \\ 
& = & S_{\max} \sum_{j=1}^N \sum_{k=1}^M G_1(p N P_j Q_k).  
\end{eqnarray}
\par
A monotonously decreasing continuous function $F(x)$ satisfies 
\begin{equation}
\label{Fx}
\sum_{k=1}^M F(k) \leq \int_1^M F(y) \ {\rm d}y + F(1),
\end{equation}
so that
\begin{equation}
\sum_{k=1}^M G_1(p N P_j Q_k) \leq \int_1^M 
 \ G_1\left(p N P_j \frac{y^{-\beta}}{\zeta_M(\beta)} 
\right) {\rm d}y 
+ G_1\left( p N P_j \frac{1}{\zeta_M(\beta)} \right)
\end{equation}
with
\begin{equation}
\zeta_M(\beta) = \sum_{\ell=1}^M \ell^{-\beta}.
\end{equation}
Then one can again use (\ref{Fx}) to obtain
\begin{equation}
\sum_{j=1}^N \sum_{k=1}^M G_1(p N P_j Q_k) \leq \sum_{\nu = 1}^4 I_{\nu},
\end{equation}
where
\begin{eqnarray} 
I_1 & = & \int_1^N {\rm d}x \int_1^M {\rm d}y 
 \ G_1\left(p N 
\frac{x^{-\alpha}}{\zeta_N(\alpha)} 
\frac{y^{-\beta}}{\zeta_M(\beta)} 
\right), \nonumber \\   
I_2 & = & \int_1^M {\rm d}y \ G_1\left( 
p N 
\frac{1}{\zeta_N(\alpha)} 
\frac{y^{-\beta}}{\zeta_M(\beta)} 
\right),
 \ \ \ I_3 = \int_1^N {\rm d}x \ G_1\left( 
p N 
\frac{x^{-\alpha}}{\zeta_N(\alpha)} 
\frac{1}{\zeta_M(\beta)} 
\right), 
\nonumber \\  I_4 & = & 
G_1\left( p N 
\frac{1}{\zeta_N(\alpha)} 
\frac{1}{\zeta_M(\beta)} 
\right).
\end{eqnarray}
\par
Using the notations  
\begin{equation}
\epsilon_1 = \frac{\sqrt{p N}}{\zeta_N(\alpha)} N^{-\alpha}, \ \ \ 
\epsilon_2 = \frac{\sqrt{p N}}{\zeta_M(\beta)} M^{-\beta}, 
\end{equation}
we see that
\begin{equation}
I_1 = \frac{MN \epsilon_1^{1/\alpha} \epsilon_2^{1/\beta}}{\alpha \beta} 
\int_{\epsilon_1}^{\epsilon_1 N^{\alpha}} {\rm d}u 
\int_{\epsilon_2}^{\epsilon_2 M^{\beta}} {\rm d}v \ \frac{G_1(uv)}{u^{1 + (1/\alpha)} v^{1 + (1/\beta)}}.
\end{equation}
Then, using the inequality
\begin{equation}
\label{g1a} 
G_1(x) \leq x^2/2, \ \ \ x \geq 0,
\end{equation}
we find
\begin{eqnarray}
\label{I1a}
I_1 & \leq & 
 \frac{MN \epsilon_1^{1/\alpha} \epsilon_2^{1/\beta}}{2 \alpha \beta} 
\int_{\epsilon_1}^{\epsilon_1 N^{\alpha}} {\rm d}u 
\int_{\epsilon_2}^{\epsilon_2 M^{\beta}} {\rm d}v \ 
u^{1 - (1/\alpha)} v^{1 - (1/\beta)} \nonumber 
\\ & = & O(N^{(2 \alpha - 1)} N^{(2 \beta - 1)}), 
\end{eqnarray}
where 
\begin{equation} N^{(2 \alpha - 1)} \equiv \left\{ \begin{array}{ll} 1, & 0 < \alpha < 1/2, \\ 
\ln N, & \alpha = 1/2, \\ N^{2 \alpha - 1}, & 1/2 < \alpha < 1. \end{array} \right. 
\end{equation}
In the case $\min(\alpha,\beta) > 1/2$, we similarly employ   
\begin{equation}
\label{g1b}
G_1(x) \leq \left\{ \begin{array}{ll} 
x^2/2, & 0 \leq x \leq 1, \\ 
x, & x \geq 1 
\end{array} \right. 
\end{equation}
to obtain
\begin{equation}
\label{I1b}
I_1 = O(N^{(\alpha,\beta)}),
\end{equation}
where
\begin{equation}
\label{Nalphabeta}
N^{(\alpha,\beta)} = \left\{ \begin{array}{ll} 
N^{(\alpha + \beta - 1)/\min(\alpha,\beta)},  & \alpha \neq \beta, \\ N^{(2 \alpha - 1)/\alpha} \ln N,  & \alpha = \beta. \end{array} \right.
\end{equation}
Using the inequality (\ref{g1a}), we can similarly derive the estimates
\begin{equation}
\label{I2I3a}
I_2 = O(N^{2 \alpha - 1} N^{(2 \beta - 1)}), \ \ \ 
I_3 = O(N^{(2 \alpha - 1)} N^{2 \beta - 1}).  
\end{equation}
If $\min(\alpha,\beta) > 1/2$, we utilise (\ref{g1b}) to find 
\begin{equation}
\label{I2I3b}
I_2 = O(N^{(\alpha + \beta - 1)/\beta}), \ \ \ 
I_3 = O(N^{(\alpha + \beta - 1)/\alpha}). \ \ \ 
\end{equation}
Moreover, one can readily see from the inequality $G_1(x) \leq x$ ($x \geq 0$) 
that
\begin{equation}
\label{I4}
I_4 = O(N^{\alpha + \beta - 1}).
\end{equation}
It follows from (\ref{I1a}), (\ref{I1b}), (\ref{I2I3a}), (\ref{I2I3b}) 
and (\ref{I4}) that    
\begin{equation}
\sum_{j=1}^N \sum_{k=1}^M G_1(p N P_j Q_k) \leq \sum_{\nu = 1}^4 I_{\nu} = o(N),
\end{equation}
which yields (\ref{2nd}).
\par
\medskip
\noindent
{\bf Step 2}
\par
\medskip
\noindent
We next prove 
\begin{equation}
\label{3rd}
\sum_{\ell=2}^{\infty} \frac{(-1)^{\ell+1}}{l} 
\sum_{j=1}^N \sum_{k=1}^M (f_{jk} S_{jk})^\ell = o(N).
\end{equation} 
Using 
\begin{equation}
G_0(x) = 1 - {\rm e}^{-x},
\end{equation}
we see that    
\begin{eqnarray}
\left| \sum_{\ell=2}^{\infty} \frac{(-1)^{\ell+1}}{l} 
\sum_{j=1}^N \sum_{k=1}^M (f_{jk} S_{jk})^\ell \right| 
& \leq & \sum_{\ell=2}^{\infty} \frac{S_{\max}^{\ell}}{\ell} \sum_{j=1}^N 
\sum_{k=1}^M f_{jk}^{\ell} 
\nonumber \\ 
& = & \sum_{\ell=2}^{\infty} \frac{S_{\max}^{\ell}}{\ell} 
\sum_{j=1}^N \sum_{k=1}^M G_0(p N P_j Q_k)^{\ell}. \nonumber \\  
\end{eqnarray}
We can again employ (\ref{Fx}) to obtain
\begin{equation}
\sum_{j=1}^N \sum_{k=1}^M G_0(p N P_j Q_k)^{\ell} \leq \sum_{\nu = 1}^4 J_{\nu},
\end{equation}
where
\begin{eqnarray} 
J_1 & = & \int_1^N {\rm d}x \int_1^M {\rm d}y 
\ G_0\left(p N 
\frac{x^{-\alpha}}{\zeta_N(\alpha)} 
\frac{y^{-\beta}}{\zeta_M(\beta)} 
\right)^{\ell}, \nonumber \\   
J_2 & = & \int_1^M {\rm d}y \ 
G_0\left( p N 
\frac{1}{\zeta_N(\alpha)} 
\frac{y^{-\beta}}{\zeta_M(\beta)} 
\right)^{\ell},
 \ \ \ 
J_3 = \int_1^N {\rm d}x \ G_0\left( p N 
\frac{x^{-\alpha}}{\zeta_N(\alpha)} 
\frac{1}{\zeta_M(\beta)} 
\right)^{\ell}, 
\nonumber \\  J_4 & = & G_0\left( p N 
\frac{1}{\zeta_N(\alpha)} 
\frac{1}{\zeta_M(\beta)} 
\right)^{\ell}.
\end{eqnarray}
Making use of the identity 
\begin{equation}
G_0(x) \leq x, \ \ \ x \geq 0,
\end{equation}
we find
\begin{equation}
\label{J1a}
J_1 = O\left( 
N^{\langle \alpha,\ell \rangle} N^{\langle \beta,\ell  \rangle} 
\right),
\end{equation}
where
\begin{equation}
N^{\langle \alpha,\ell \rangle} = \left\{ \begin{array}{ll} 
N^{(2 - \ell)/2}, & \ell < 1/\alpha, \\ 
N^{(2 \alpha  - 1)/(2 \alpha)} \ln N,  &  \ell = 1/\alpha, \\ 
N^{\ell (2 \alpha - 1)/2}, & \ell > 1/\alpha.
\end{array} \right.
\end{equation}
In the case $\alpha + \beta > 1$, by means of
\begin{equation}
\label{g0}
G_0(x) \leq \left\{ \begin{array}{ll} x, & 0 \leq x \leq 1, \\ 
1, & x \geq 1, \end{array} \right.
\end{equation}
we obtain 
\begin{equation}
\label{J1b}
J_1 = \left\{ \begin{array}{ll} 
O(N^{(\alpha,\beta)}), & \alpha > 1/2, \ \beta > 1/2, \\ 
O(N^{2 \alpha - 1} \ln N), & \alpha > 1/2, \ \beta \leq 1/2, \ 
\ell = 1/\beta, \\     
O(N^{2 \alpha - 1}), & \alpha > 1/2, \ \beta \leq 1/2, \ 
\ell \neq 1/\beta, \\    
O(N^{2 \beta - 1} \ln N), & \alpha \leq 1/2, \ \beta > 1/2, \ 
\ell = 1/\alpha, \\  
O(N^{2 \beta - 1}), & \alpha \leq 1/2, \ \beta > 1/2, \ 
\ell \neq 1/\alpha. 
\end{array} \right.
\end{equation}
Here the symbol $N^{(\alpha,\beta)}$ is defined in (\ref{Nalphabeta}). 
The inequality (\ref{g0}) similarly gives the estimates 
\begin{eqnarray}
\label{J2J3}
J_2 & = & \left\{ \begin{array}{ll} 
O(N^{1 + \ell (\alpha - 1)}), & \ell < 1/\beta, \\       
O(N^{(\alpha + \beta - 1)/\beta} \ln N), & \ell = 1/\beta, \\ 
O(N^{(\alpha + \beta - 1)/\beta}), & \ell > 1/\beta,
\end{array} \right. \nonumber \\ 
J_3 & = & \left\{ \begin{array}{ll} 
O(N^{1 + \ell (\beta - 1)}), & \ell < 1/\alpha, \\       
O(N^{(\alpha + \beta - 1)/\alpha} \ln N), & \ell = 1/\alpha, \\ 
O(N^{(\alpha + \beta - 1)/\alpha}), & \ell > 1/\alpha.
\end{array} \right. 
\end{eqnarray}
Moreover it is evident from the  inequality $G_0(x) \leq 1$ 
($x \geq 0$) that
\begin{equation}
\label{J4}
J_4 = O(1).
\end{equation} 
Now we can easily see from (\ref{J1a}), (\ref{J1b}), (\ref{J2J3}) and 
(\ref{J4}) that 
\begin{equation}
\sum_{j=1}^N \sum_{k=1}^M G_0(p N P_j Q_k)^{\ell} \leq \sum_{\nu = 1}^4 
J_{\nu} = o(N) 
\end{equation}
for any $\ell \geq 2$. This relation results in the asymptotic 
estimate (\ref{3rd}).

\section*{Appendix B}
\setcounter{equation}{0}
\renewcommand{\theequation}{B.\arabic{equation}}

Let us first discuss the spectral density of the adjacency matrix ${\cal A}$. 
The eigenvalues $\mu_j$, $j = 1,2,\cdots,N + M$ of ${\cal A}$ consist of 
$M$ pairs $\pm \nu_j$, $j = 1,2,\cdots,M$ and $N - M$ zeros. Note that 
$\nu_j^2 > 0$ are identified with the eigenvalues of the $M \times M$ 
correlation matrix $V$ with
\begin{equation}
V_{kl} = \sum_{j=1}^N C_{jk} C_{jl}.
\end{equation}
\par
Using the notations 
\begin{equation}
\phi_j = \Phi_j, \ \ \ j = 1,2,\cdots,N
\end{equation}
and
\begin{equation}
\psi_k = \Phi_{k+N}, \ \ \ k = 1,2,\cdots,M,
\end{equation}
we can rewrite the partition function $Z$ defined in (\ref{pf}) as      
\begin{equation}
Z(\mu) = 
\int \prod_{j=1}^N {\rm d}\phi_j 
\int \prod_{k=1}^M {\rm d}\psi_k
\ {\rm exp}\left( 
\frac{i}{2} \mu \sum_{j=1}^N \phi_j^2 
+ \frac{i}{2} \mu \sum_{k=1}^M \psi_k^2 
- i \sum_{j=1}^N \sum_{k=1}^M C_{jk} \phi_j \psi_k 
\right).
\end{equation}
Then we introduce the replica variables 
\begin{equation}
{\vec \phi}_j = (\phi_j^{(1)},\phi_j^{(2)},\cdots,\phi_j^{(n)}), \ \ \ 
{\vec \psi}_k = (\psi_k^{(1)},\psi_k^{(2)},\cdots,\psi_k^{(n)}) 
\end{equation}
and
\begin{equation}  
{\rm d}{\vec \phi}_j = {\rm d}\phi_j^{(1)}{\rm d}\phi_j^{(2)} 
\cdots {\rm d}\phi_j^{(n)}, \ \ \  
{\rm d}{\vec \psi}_k= {\rm d}\psi_k^{(1)}{\rm d}\psi_k^{(2)} 
\cdots {\rm d}\psi_k^{(n)}   
\end{equation}
to obtain
\begin{eqnarray}
\langle Z^n \rangle & = &  
\int \prod_{j=1}^N {\rm d}{\vec \phi}_j 
\int \prod_{k=1}^M {\rm d}{\vec \psi}_k
\ {\rm exp}\left( 
\frac{i}{2} \mu \sum_{j=1}^N {\vec \phi}_j^2 
+ \frac{i}{2} \mu \sum_{k=1}^M {\vec \psi}_k^2 \right) 
\nonumber \\ & & \times  
\left\langle {\rm exp}\left(   
- i \sum_{j=1}^N \sum_{k=1}^M C_{jk} {\vec \phi}_j \cdot {\vec \psi}_k  
\right) \right\rangle.
\end{eqnarray}
Now we can see from (\ref{asymptotic}) that
\begin{equation} 
\left\langle {\rm exp}\left( - i \sum_{j=1}^N \sum_{k=1}^M C_{jk} 
{\vec \phi}_j \cdot {\vec \psi}_k \right) \right\rangle \sim  
{\rm exp}\left\{ p N \sum_{j=1}^N \sum_{k=1}^M 
P_j Q_k \left( {\rm e}^{-i {\vec \phi}_j \cdot {\vec \psi}_k} - 1 \right) 
\right\}. 
\end{equation}
It should be noted that this asymptotic relation holds if ${\vec \phi}_j 
\cdot {\vec \psi}_k$ is in the neighbourhood of the origin. This condition 
is justified in the limit of large mean degree $p \rightarrow \infty$, 
since ${\vec \phi}_j^2$ and ${\vec \psi}_k^2$ are scaled as $O(p^{-1/2})$ or 
$O(p^{-1})$ (see eqs. (\ref{scale1}) and (\ref{scale2})).  
\par
Using the notation   
\begin{equation}
{\tilde \xi}_j({\vec \phi}) = \delta({\vec \phi}- {\vec \phi}_j), \ \ \ 
{\tilde \eta}_k({\vec \psi}) = \delta({\vec \psi}- {\vec \psi}_k),  
\end{equation}
we obtain 
\begin{eqnarray} 
& & \left\langle {\rm exp}\left( - i \sum_{j=1}^N \sum_{k=1}^M 
C_{jk} {\vec \phi}_j \cdot {\vec \psi}_k \right) \right\rangle 
\nonumber \\ & \sim & {\rm exp}\left\{ 
p N \sum_{j=1}^N \sum_{k=1}^M P_j Q_k \int {\rm d}{\vec \phi} 
\int {\rm d}{\vec \psi} \ 
{\tilde \xi}_j({\vec \phi}) {\tilde \eta}_k({\vec \psi}) \left( 
{\rm e}^{- i {\vec \phi} \cdot {\vec \psi}}  
- 1 \right) \right\}, \nonumber \\  
\end{eqnarray}
so that we find
\begin{eqnarray}
\label{znaverage}
& & \langle Z^n \rangle \sim \int \prod_{j=1}^N {\rm d}{\vec \phi}_j 
\int \prod_{k=1}^M {\rm d}{\vec \psi}_k  
\nonumber \\ & & \times \ {\rm exp}\left\{ 
\frac{i}{2} \mu \sum_{j=1}^N \int {\rm d}{\vec \phi} \  
{\tilde \xi}_j({\vec \phi}) {\vec \phi}^2 +  
\frac{i}{2} \mu \sum_{k=1}^M \int {\rm d}{\vec \psi} \  
{\tilde \eta}_k({\vec \psi}) {\vec \psi}^2 \right\} 
\nonumber \\ 
& & \times \ {\rm exp}\left[ 
 p N \sum_{j=1}^N \sum_{k=1}^M P_j Q_k  
\int {\rm d}{\vec \phi} \int {\rm d}{\vec \psi} \  
{\tilde \xi}_j({\vec \phi}) {\tilde \eta}_k({\vec \psi}) \left( {
\rm e}^{- i {\vec \phi} \cdot {\vec \psi}} 
- 1 \right) \right] 
\nonumber \\ & = & 
\int \prod_{j=1}^N {\rm d}{\vec \phi}_j 
\int \prod_{k=1}^M {\rm d}{\vec \psi}_k 
\int \prod_{j=1}^N {\cal D}\xi_j({\vec \phi}) 
\int \prod_{k=1}^M {\cal D}\eta_k({\vec \psi}) 
\nonumber \\ & & \times  
\prod_{j=1}^N \prod_{\vec \phi} 
\delta(\xi_j({\vec \phi}) - {\tilde \xi}_j({\vec \phi})) 
\prod_{k=1}^M \prod_{\vec \psi} 
\delta(\eta_k({\vec \psi}) - {\tilde \eta}_k({\vec \psi})) 
 \ {\rm e}^{S_1 + S_2}.
\end{eqnarray}
Here $S_1$ and $S_2$ are defined in (\ref{s1}) and (\ref{s2}). 
The auxiliary functions $\xi_j({\vec \phi})$ and 
$\eta_k({\vec \psi})$ satisfy (\ref{xieta}). If $J$ is the 
Laplacian matrix ${\cal L}$, we can similarly derive the 
same formula (\ref{znaverage}) for $\langle Z^n \rangle$, except the 
change of $S_2$ according to (\ref{fpp}).
\par
It can readily be seen that
\begin{eqnarray}
& & \int \prod_{j=1}^N {\rm d}{\vec \phi}_j \prod_{j=1}^N \prod_{\vec \phi} 
\delta(\xi_j({\vec \phi}) - {\tilde \xi}_j({\vec \phi})) \nonumber \\ 
& = &  \int \prod_{j=1}^N {\rm d}{\vec \phi}_j 
\int \prod_{j=1}^N {\cal D}a_j({\vec \phi}) \ {\rm exp} \left[
2 \pi i \sum_{j=1}^N \int {\rm d}{\vec \phi} \ a_j({\vec \phi})
\left\{ \xi_j({\vec \phi})-{\tilde \xi}_j({\vec \phi}) \right\} \right]
\nonumber \\ 
& = &  \int \prod_{j=1}^N {\cal D}a_j({\vec \phi}) \ {\rm exp}\left[ 
\sum_{j=1}^N \left\{ 2 \pi i \int 
{\rm d}{\vec \phi} \ a_j({\vec \phi}) \xi_j({\vec \phi}) - W_j \right\} \right], 
\end{eqnarray}
where
\begin{eqnarray}
W_j & = & - \ln \int {\rm d}{\vec \phi}_j \ {\rm exp}\left\{ 
- 2 \pi i \int {\rm d}{\vec \phi} \ a_j({\vec \phi}) 
{\tilde \xi}_j({\vec \phi}) \right\} \nonumber \\ 
& = &  - \ln \int {\rm d}{\vec \phi}_j \ {\rm exp}\left\{ 
- 2 \pi i a_j({\vec \phi}_j) \right\}. 
\end{eqnarray}
In the limit $N \rightarrow \infty$, the dominant contribution comes from 
the stationary point satisfying 
\begin{equation}
\frac{\delta}{\delta a_j({\vec \phi})} \left\{  
2 \pi i \int {\rm d}{\vec \phi} \ a_j({\vec \phi}) \xi_j({\vec \phi}) 
- W_j \right\} = 2 \pi i \xi_j({\vec \phi}) - 2 \pi i {\rm e}^{ 
- 2 \pi i a_j({\vec \phi}) + W_j } = 0,
\end{equation}
which means  
\begin{equation}
- \int {\rm d}{\vec \phi} \ \xi_j({\vec \phi}) \ln \xi_j({\vec \phi}) = 2 \pi i
\int {\rm d}{\vec \phi} \ a_j({\vec \phi}) \xi_j({\vec \phi}) - W_j.
\end{equation}
Therefore we find an asymptotic estimate
\begin{equation} 
\int \prod_{j=1}^N {\rm d}{\vec \phi}_j \prod_{j=1}^N \prod_{\vec \phi} 
\delta(\xi_j({\vec \phi}) - {\tilde \xi}_j({\vec \phi})) \nonumber \\ 
\sim {\rm exp}\left\{ -  
\sum_{j=1}^N \int {\rm d}{\vec \phi} \ 
\xi_j({\vec \phi}) \ln \xi_j({\vec \phi}) 
\right\}.  
\end{equation}
One can similarly derive another estimate
\begin{equation} 
\int \prod_{k=1}^M {\rm d}{\vec \psi}_k \prod_{k=1}^M \prod_{\vec \psi} 
\delta(\eta_k({\vec \psi}) - {\tilde \eta}_k({\vec \psi})) \nonumber \\ 
\sim {\rm exp}\left\{ -  
\sum_{k=1}^M \int {\rm d}{\vec \psi} \ \eta_k({\vec \psi}) \ln \eta_k({\vec \psi}) 
\right\}  
\end{equation}
in the limit $M \rightarrow \infty$. Then we arrive at  
\begin{equation}
\langle Z^n \rangle \sim \int 
\prod_{j=1}^N {\cal D}\xi_j({\vec \phi}) 
\prod_{k=1}^M {\cal D}\eta_k({\vec \psi}) 
\ {\rm e}^{S_0 + S_1 + S_2}, 
\end{equation}
where $S_0$ is defined in (\ref{s0}).

\section*{Appendix C}
\setcounter{equation}{0}
\renewcommand{\theequation}{C.\arabic{equation}}

Putting the Gaussian ansatz (\ref{gaussian}) into (\ref{variational}), 
we see in the limit $n \rightarrow 0$ that   
\begin{eqnarray}
\xi_j({\vec \phi}) & = & \Theta_j \ {\rm exp}\left[ \frac{i}{2} \mu {\vec \phi}^2 + 
p N P_j \sum_{k=1}^M Q_k \left\{  g_k({\vec \phi}) - 1 \right\}  
\right], \nonumber \\ 
\eta_k({\vec \psi}) & = & \Omega_k \ {\rm exp}\left[ \frac{i}{2} \mu {\vec \psi}^2 + 
p N Q_k \sum_{j=1}^N P_j \left\{ h_j({\vec \psi})- 1 \right\} \right], 
\end{eqnarray}
where
\begin{equation}
g_k({\vec \phi}) = 
\left\{ \begin{array}{ll} 
\displaystyle {\rm exp}\left( - \frac{i \tau_k}{2} {\vec \phi}^2 \right), 
& {\rm if} \ J \ {\rm is} \ {\rm the} \ {\rm adjacency} \ {\rm matrix} 
\ {\cal A}, \\ 
\displaystyle {\rm exp}\left( - \frac{i}{2 (1 - \tau_k)} {\vec \phi}^2 \right), 
& {\rm if} \ J \ {\rm is} \ {\rm the} \ {\rm Laplacian} \ {\rm  matrix} 
\ {\cal L} 
\end{array} \right.
\end{equation}
and 
\begin{equation}
h_j({\vec \psi}) = 
\left\{ \begin{array}{ll} 
\displaystyle {\rm exp}\left( - \frac{i \sigma_j}{2} {\vec \psi}^2 \right), 
& {\rm if} \ J \ {\rm is} \ {\rm the} \ {\rm adjacency} \ {\rm matrix} 
\ {\cal A}, \\ 
\displaystyle {\rm exp}\left( - \frac{i}{2 (1 - \sigma_j)} {\vec \psi}^2 \right), 
& {\rm if} \ J \ {\rm is} \ {\rm the} \ {\rm Laplacian} \ {\rm  matrix} 
\ {\cal L}.
\end{array} \right.
\end{equation}
\par
Let us first consider the adjacency matrix ${\cal A}$. We are in a position to 
take the limit $p \rightarrow \infty$ with 
the scalings
\begin{equation}
\label{scale1}
\mu = O(p^{1/2}), \ \ \ {\vec \phi}^2 = O(p^{-1/2}), \ \ \  \sigma_j = O(p^{-1/2}), \ \ \ 
{\vec \psi}^2 = O(p^{-1/2}), \ \ \  \tau_k = O(p^{-1/2}).
\end{equation}
Then we obtain
\begin{equation}
\label{sigmatau}
\mu - \frac{1}{\sigma_j} - p N P_j \sum_{k=1}^M Q_k \tau_k = 0, \ \ \ 
\mu - \frac{1}{\tau_k} - p N Q_k \sum_{j=1}^N P_j \sigma_j = 0. 
\end{equation}
The variational equations (\ref{variational}) are satisfied by the 
Gaussian ansatz (\ref{gaussian}), if $\sigma_j$ and $\tau_k$ are 
determined by these equations.  
\par
In order to analytically treat (\ref{sigmatau}), we define the scaling variables
\begin{equation}
E = \mu/\sqrt{p}, \ \ \ s(x) = \sqrt{p} \ \sigma_j, \ \ \ x = j/N, \ \ \  
t(y) = \sqrt{p} \ \tau_k, \ \ \  y = k/M.
\end{equation}
Then it is straightforward to find
\begin{equation}
\frac{x^{\alpha}}{s(x)} = E x^{\alpha} - (1 - \alpha) (1 - \beta) 
\int_0^1 y^{-\beta} t(y) \ {\rm d}y
\end{equation}
and
\begin{equation}
\frac{y^{\beta}}{t(y)} = E y^{\beta} - c (1 - \alpha) (1 - \beta) 
\int_0^1 x^{-\alpha} s(x) \ {\rm d}x
\end{equation}
in the limit (\ref{limit}). Using the notations
\begin{equation}
S = c (1 - \alpha) (1 - \beta) \int_0^1 x^{-\alpha} s(x) \ {\rm d}x, \ \ \  
T =  (1 - \alpha) (1 - \beta) \int_0^1 y^{-\beta} t(y) \ {\rm d}y,
\end{equation}
we obtain
\begin{equation}
\frac{S}{c (1 - \alpha) (1 - \beta)} = \int_0^1 \frac{1}{E x^{\alpha} - T} 
{\rm d}x, \ \ \ 
\frac{T}{(1 - \alpha) (1 - \beta)} = \int_0^1 \frac{1}{E y^{\beta} - S} 
{\rm d}y. 
\end{equation}
\par
In order to evaluate the behaviour of $S$ and $T$ in the tail region $E 
\rightarrow \infty$, we write
\begin{equation}
S = E (s^{(R)} - i s^{(I)}), \ \ \    
T = E (t^{(R)} - i t^{(I)})
\end{equation}
with real $s^{(R)}$, $s^{(I)}$, $t^{(R)}$ and $t^{(I)}$. Then it can 
be seen that
\begin{eqnarray}
& & \frac{E (s^{(R)} - i s^{(I)})}{c (1 - \alpha) (1 - \beta)} 
 = \frac{1}{E} 
\int_0^1 \frac{1}{x^{\alpha} - t^{(R)} + i t^{(I)}} {\rm d}x
 = \frac{1}{\alpha E} 
\int_0^1 \frac{s^{(1 - \alpha)/\alpha}}{s - t^{(R)} + i t^{(I)}} {\rm d}s
\nonumber \\ & = & \frac{1}{\alpha E} 
\int_0^1 \frac{s^{(1 - \alpha)/\alpha} \ (s - t^{(R)})}{(s - t^{(R)})^2 
+(t^{(I)})^2} {\rm d}s
- \frac{i}{\alpha E} 
\int_0^1 \frac{s^{(1 - \alpha)/\alpha} \ t^{(I)}}{(s - t^{(R)})^2 
+(t^{(I)})^2} {\rm d}s.
\end{eqnarray}
Let us employ an asymptotic formula\cite{RAKK} 
\begin{equation}
\frac{\epsilon}{(u - a)^2 + \epsilon^2} \sim \pi \delta(u - a), \ \ \ 
\epsilon \downarrow 0 
\end{equation}
and obtain an estimate 
\begin{eqnarray}
\frac{E (s^{(R)} - i s^{(I)})}{c (1 - \alpha) (1 - \beta)} 
& \sim & \frac{1}{\alpha E} \int_0^1 s^{(1/\alpha)-2} 
{\rm d}s - \frac{i \pi}{\alpha E} \int_0^1 s^{(1 - \alpha)/\alpha} 
\delta(s - t^{(R)}) {\rm d}s 
\nonumber \\ 
& = & 
\frac{1}{E (1 - \alpha)} - \frac{i \pi}{\alpha E} 
(t^{(R)})^{(1 - \alpha)/\alpha}, \ \ \ E \rightarrow \infty,
\end{eqnarray}
so that
\begin{equation}
\label{srsi}
s^{(R)} \sim \frac{c (1 - \beta)}{E^2}, \ \ \ 
s^{(I)} \sim \frac{c (1 - \alpha)(1 - \beta)\pi}{\alpha E^2} 
(t^{(R)})^{(1 - \alpha)/\alpha}.
\end{equation}
One can similarly derive  
\begin{eqnarray}
& & \frac{E (t^{(R)} - i t^{(I)})}{(1 - \alpha) (1 - \beta)} 
 = \frac{1}{E} 
\int_0^1 \frac{1}{y^{\beta} - s^{(R)} + i s^{(I)}} {\rm d}x
 = \frac{1}{\beta E} 
\int_0^1 \frac{t^{(1 - \beta)/\beta}}{t - s^{(R)} + i s^{(I)}} {\rm d}t
\nonumber \\ 
& \sim &  
\frac{1}{E (1 - \beta)} - \frac{i \pi}{\beta E} 
(s^{(R)})^{(1 - \beta)/\beta}, \ \ \ E \rightarrow \infty, 
\end{eqnarray}
so that
\begin{equation}
\label{trti}
t^{(R)} \sim \frac{1 - \alpha}{E^2}, \ \ \ 
t^{(I)} \sim \frac{(1 - \alpha)(1 - \beta)\pi}{\beta E^2} 
(s^{(R)})^{(1 - \beta)/\beta}.
\end{equation}
It follows from (\ref{srsi}) and ({\ref{trti}) that
\begin{equation}
\label{siti}
s^{(I)} \sim \frac{1 - \beta}{\alpha} \pi c 
\left( \frac{1 - \alpha}{E^2} \right)^{1/\alpha}, 
\ \ \ t^{(I)} \sim \frac{1 - \alpha}{\beta} \pi c^{(1 - \beta)/\beta}  
\left( \frac{1 - \beta}{E^2} \right)^{1/\beta}. 
\end{equation}
\par
Now we can evaluate the asymptotic behaviour of the spectral 
density $\rho(\mu)$ in the tail region $E \rightarrow \infty$. 
Eqs. (\ref{rhomu}) and (\ref{s0s1s2}) can be utilised as  
\begin{eqnarray}
\rho(\mu) & = & 
 \lim_{n \rightarrow 0} \frac{2}{(N+M) n \pi} {\rm Im} 
\frac{\partial}{\partial \mu}(S_0 + S_1 + S_2) 
\nonumber \\ & = & \lim_{n \rightarrow 0} \frac{1}{(N+M) n \pi} {\rm Re} 
\left( \sum_{j=1}^N \int {\rm d}{\vec \phi} 
\ \xi_j({\vec \phi}) {\vec \phi}^2 + 
\sum_{k=1}^M \int {\rm d}{\vec \psi} 
\ \eta_k({\vec \psi}) {\vec \psi}^2 \right)
\nonumber \\ & = & - \frac{1}{(N+M) \pi} {\rm Im} 
\left( \sum_{j=1}^N \sigma_j  + \sum_{k=1}^M \tau_k 
\right)
\nonumber \\ & \sim & - \frac{1}{\sqrt{p} (1 + c) \pi} {\rm Im} 
\left(c \int_0^1 s(x) {\rm d}x  + \int_0^1 t(y) {\rm d}y   
\right)
\end{eqnarray}
in the limit (\ref{limit}). Here
\begin{eqnarray}
\int_0^1 s(x) {\rm d}x & = & \int_0^1 \frac{1}{E - T x^{-\alpha}} {\rm d}x 
= \frac{1}{E} + \frac{T}{E} \int_0^1 \frac{1}{E x^{\alpha} - T} {\rm d}x
\nonumber \\ 
& = & \frac{1}{E} + \frac{ST}{c E (1 - \alpha) (1 - \beta)}
\end{eqnarray}
and we can similarly obtain 
\begin{equation}
\int_0^1 t(y) {\rm d}y 
= \frac{1}{E} + \frac{ST}{E (1 - \alpha) (1 - \beta)}.
\end{equation}
Then it can be seen from (\ref{srsi}), (\ref{trti}) and (\ref{siti}) that 
\begin{eqnarray}
\rho(\mu) & \sim & - \frac{2}{\sqrt{p} E (1 + c) \pi 
(1 - \alpha) (1 - \beta)} {\rm Im}( ST )
\nonumber \\  & = & \frac{2 E}{\sqrt{p}(1 + c) \pi 
(1 - \alpha) (1 - \beta)} \left\{ s^{(R)} t^{(I)} +  s^{(I)} t^{(R)} \right\} 
\nonumber \\ 
& \sim & \frac{2}{\sqrt{p} (1 + c)} \left\{ 
\frac{c^{1/\beta} (1 - \beta)^{1/\beta}}{\beta E^{(2/\beta)+ 1}}   
+ \frac{c (1 - \alpha)^{1/\alpha}}{\alpha E^{(2/\alpha)+ 1}} \right\}.
\end{eqnarray} 
This gives the asymptotic spectral density of the adjacency matrix ${\cal A}$ 
in the tail region $E \rightarrow \infty$. 
\par
We next compute the spectral density of the Laplacian matrix ${\cal L}$. 
Using the scalings
\begin{equation}
\label{scale2}
\mu = O(p), \ \ \ {\vec \phi}^2 = O(p^{-1}), \ \ \  \sigma_j = O(p^{-1}), \ \ \ 
{\vec \psi}^2 = O(p^{-1}), \ \ \  \tau_k = O(p^{-1})
\end{equation}
and taking the limit $p \rightarrow \infty$, we find 
\begin{equation}
\mu - \frac{1}{\sigma_j} - p N P_j  = 0, \ \ \ 
\mu - \frac{1}{\tau_k} - p N Q_k = 0, 
\end{equation}
so that
\begin{eqnarray}
{\rm Im} \sigma_j & = & {\rm Im}\frac{1}{\mu + i \epsilon - p N P_j} = 
- \pi \delta( \mu - p N P_j), \nonumber \\  
{\rm Im} \tau_k & = & {\rm Im}\frac{1}{\mu + i \epsilon - p N Q_k} = 
- \pi \delta( \mu - p N Q_k),
\end{eqnarray} 
where $\epsilon$ is an infinitesimal positive number. Then it follows 
in the limit (\ref{limit}) that
\begin{eqnarray}
{\rm Im}\sum_{j=1}^N \sigma_j & \sim & - N \pi \int_0^1 {\rm d}x \ 
\delta(\mu - p (1 - \alpha) x^{-\alpha}) 
\nonumber \\ & = &  
 - N \pi \frac{\{p (1 - \alpha)\}^{1/\alpha}}{\alpha} 
\frac{1}{\mu^{(1/\alpha)+1}} \ H\left\{ \mu - p (1 - \alpha) \right\} 
\end{eqnarray}   
and
\begin{eqnarray}
{\rm Im}\sum_{k=1}^M \tau_k & \sim & - M \pi \int_0^1 {\rm d}y \ 
\delta(\mu - p c (1 - \beta) y^{-\beta}) 
\nonumber \\ & = &  
 - M \pi \frac{\{p c(1 - \beta)\}^{1/\beta}}{\beta} 
\frac{1}{\mu^{(1/\beta)+1}} \ H\left\{ \mu - p c (1 - \beta) \right\}, 
\end{eqnarray}
where $H(x)$ is defined in (\ref{hx}). Therefore we arrive at
\begin{eqnarray}
\rho(\mu) &  = & - \frac{1}{(N+M) \pi} {\rm Im} 
\left( \sum_{j=1}^N \sigma_j  + \sum_{k=1}^M \tau_k 
\right)\nonumber \\ 
& \sim & 
\frac{c \{p (1 - \alpha)\}^{1/\alpha}}{(1 + c) \alpha} 
\frac{1}{\mu^{(1/\alpha)+1}} \ H\left\{ \mu - p (1 - \alpha) \right\}  
\nonumber \\ 
& & + \frac{\{p c (1 - \beta)\}^{1/\beta}}{(1 + c) \beta} 
\frac{1}{\mu^{(1/\beta)+1}} \ H\left\{ \mu - p c (1 - \beta) \right\}.
\end{eqnarray}
This gives the asymptotic spectral density of the Laplacian matrix 
${\cal L}$ in the region $\mu = O(p)$.

\end{document}